\documentclass[10pt,letterpaper]{article}
\usepackage{opex3}
\usepackage{amssymb,amsmath}
\usepackage{cite}

\begin{document}

\title{Polarizability of nanowires at surfaces: Exact solution for general geometry}

\author{Jesper Jung$^{*}$ and Thomas G. Pedersen}

\address{Department of Physics and Nanotechnology, Aalborg
University, Skjernvej 4A, DK-9220 Aalborg {\O}st, Denmark and
Interdisciplinary Nanoscience Center (iNANO), Denmark}

\email{$^{*}$jung@nano.aau.dk} 



\begin{abstract}
The polarizability of a nanostructure is an important parameter that
determines the optical properties. An exact semi-analytical solution
of the electrostatic polarizability of a general geometry consisting
of two segments forming a cylinder that can be arbitrarily buried in
a substrate is derived using bipolar coordinates, cosine-, and
sine-transformations. Based on the presented expressions, we analyze
the polarizability of several metal nanowire geometries that are
important within plasmonics. Our results provide physical insight
into the interplay between the multiple resonances found in the
polarizability of metal nanowires at surfaces.
\end{abstract}



\section{Introduction}
The electric polarizability of a nanoparticle, i.e. the relative
tendency of the electron cloud to be distorted from its normal shape
by an external field, is an important concept because it determines
the particle's optical properties. Given the polarizability, it is
usually straightforward to determine the light scattering and
absorption properties of a nanoparticle
\cite{maier2007,novotny2006}. By illuminating a \emph{metal}
nanoparticle, collective excitations of the free conduction
electrons of the metal can be resonantly excited. Such excitations
are known as \emph{particle plasmons} or \emph{localized surface
plasmons}\cite{Zayats2003,Maier2005,Murray2007,lal2007} and give
rise to resonances in the polarizability. Thus, given the
polarizability of a metal nanoparticle, its particle plasmons can
easily be identified. Particle plasmon resonances are interesting
from an application point of view because they allow for strong
light scattering and large electromagnetic fields in the vicinity of
the particles. These properties suggest a variety of applications,
e.g. within surface enhanced raman scattering \cite{moskovits1985},
biomedical detection \cite{cao2002,haes2002}, and plasmon enhanced
solar cells \cite{catchpole2008,atwater2010,ferry2010}. Spherical
nanoparticles can be analyzed analytically using the old, but famous
Lorenz-Mie scattering theory \cite{Lorenz1890,Mie1908,bohren1983}
and, also, the ''two-dimensional'' case of a small cylinder has been
analyzed a long time ago \cite{Rayleigh1918,Wait1955}. However, for
more complicated structures, no analytical solutions can in general
be found. In the analysis of such structures, sophisticated
numerical schemes such as finite difference time domain
\cite{taflove2000}, finite element\cite{jin2002}, or Green's
function integral
equation\cite{søndergaard2007,martin1995,abajo2002,jung2010}
approaches are often utilized. However, such approaches are
complicated and from a computational point of view often very
time-consuming. Thus, analytical modeling, whenever possible, is
always favorable.

\begin{figure}[h]
  \begin{center}
  \includegraphics[width=\textwidth]{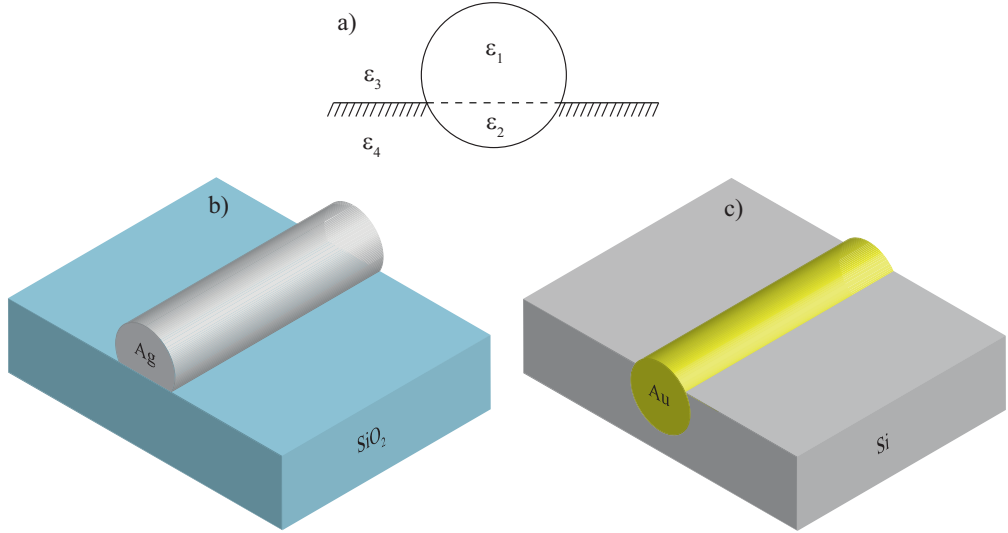}
  \caption{(a) cross section of the geometry under consideration.
  The optical properties are described by the four dielectric constants $\varepsilon_1$ to $\varepsilon_4$. (b) and (c) two examples of geometries that can be analyzed
  using the approach.}
  \label{fig1:structure2d3d}
\end{center}
\end{figure}

In the present paper, we present a semi-analytical analysis of the
polarizability of the very general geometry shown in
Fig.~\ref{fig1:structure2d3d}(a).
The geometry is two-dimensional in the sense that it is invariant
under translation along the direction out of the paper. It consists
of two domains, described by the dielectric constants
$\varepsilon_1$ and $\varepsilon_2$, which together form a cylinder.
This cylinder may be placed in a background composed of two
semi-infinite half-spaces with optical properties described by the
dielectric constants $\varepsilon_3$ and $\varepsilon_4$. As the
interface can cut the cylinder at an arbitrary position and all four
dielectric constants can be chosen freely, the structure allows for
the study of several different geometries. Two examples are depicted
in Fig.~\ref{fig1:structure2d3d}(b) and (c). In (b), we have chosen
$\varepsilon_1=\varepsilon_{\text{Ag}}$,
$\varepsilon_2=\varepsilon_4=\varepsilon_{\text{SiO}_2}$, and
$\varepsilon_{3} = \varepsilon_{\text{air}}$, whereas in (c), we
have chosen $\varepsilon_1 = \varepsilon_2 =
\varepsilon_{\text{Au}}$, $\varepsilon_4 = \varepsilon_{\text{Si}}$,
and $\varepsilon_3 = \varepsilon_{\text{air}}$. Both configurations
are interesting in connection with plasmon assisted solar cells, see
e.g. Refs. \cite{hallermann2008,yoon2011}. Previously, simpler
geometries such as half cylinders \cite{waterman1999} or double
half-cylinders in \emph{homogenous} surroundings have been analyzed
\cite{radchik1994,salandrino2007,pitkonen2010}. Also, a double
half-sphere consisting of two conjoint hemispheres with different
dielectric constants has been analyzed \cite{kettunen2007}. Recently
convex and concave rough metal surfaces formed by cylindrical
protrusions have been studied by Pendry and coworkers using the
approach of transformation optics (TO) \cite{lou2010,lou2011}. They
analyzed cut cylindrical protrusions on semi-infinite metal
surfaces. This is opposite to the present work, where we only
consider dielectric surfaces. In fact, the present approach cannot
be used to analyze metal surfaces because we are interested in the
polarizability which will lose its well defined meaning for a metal
nanowire on an infinite metal surface. However, the TO approach of
Refs.~\cite{lou2010,lou2011} is also restricted in the sense that
the dielectric constant of the protrusion and the surface must be
identical. Thus, by carefully comparing the TO approach to the
present it can be seen that they are in fact capable of analyzing
complimentary structures. The present paper is a significant
extension of our previous work, presented in Ref. \cite{jung2011},
where we analyzed the polarizability of two conjoint half-cylinders
embedded in two semi-infinite half spaces. Compared to
Ref.~\cite{jung2011}, the present approach is much more general in
that we allow for an arbitrary degree of burial of the nanowire. The
approach of Ref.~\cite{jung2011} is restricted to half-buried
nanowires only. As a consequence the new scheme is better suited for
representing realistic experimental geometries. It can, e.g., handle
nanowires with very different aspect ratios, i.e. wires with very
different height/width ratios, and it can handle arbitrary contact
angles between a nanowire and the supporting substrate. Furthermore,
the present work provides more physical insight into the different
resonances that come into play when the polarizability of
complicated metal nanostructures are considered.

The paper is organized as follows. In Sec.~\ref{sec:theory}, we
present a derivation of the electrostatic polarizability of the
geometry sketched in Fig.~\ref{fig1:structure2d3d}(a).
Section~\ref{sec:results} presents calculations and an analysis of
the polarizability and its resonances for several important
geometries. In Sec.~\ref{sec:conclusion}, we offer our conclusions.

\section{Theory}\label{sec:theory}
\begin{figure}[h]
  \begin{center}
  \includegraphics[width=6cm]{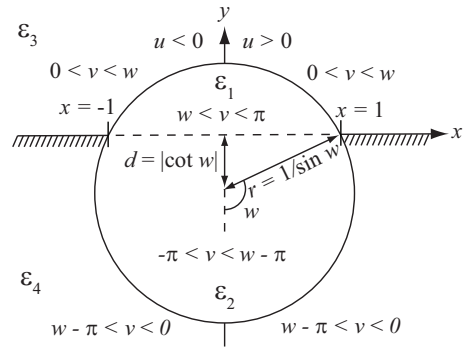}
  \caption{Cross section of the geometry analyzed. In bipolar coordinates ($u$ and $v$), the four different regions in the $xy$ plane
  are given as $\varepsilon_1:\left\{-\infty<u<\infty \ \text{and} \ w<v<\pi \right\}$, $\varepsilon_2:\left\{-\infty<u<\infty \ \text{and} \ -\pi<v<w-\pi \right\}$,
  $\varepsilon_3:\left\{-\infty<u<\infty \ \text{and} \ 0<v<w \right\}$, and $\varepsilon_4:\left\{-\infty<u<\infty \ \text{and} \ w-\pi<v<0 \right\}$.}
  \label{fig2:structure2d}
\end{center}
\end{figure}

A detailed schematic of the geometry studied is shown in
Fig.~\ref{fig2:structure2d}. It consists of four regions 1, 2, 3,
and 4. A cylinder of radius $r$ is cut by an infinite interface at
an arbitrary position $d$ from its center $|d|\leq r$. The segment
of the cylinder above (below) the interface is denoted 1 (2) and has
optical properties described by $\varepsilon_1$ ($\varepsilon_2$).
The surrounding medium above (below) the interface is denoted 3 (4)
and has a dielectric constant $\varepsilon_3$ ($\varepsilon_4$).
To comply with the symmetry of the problem it is an advantage to
switch to the bipolar coordinates $u$ and $v$ defined as
\cite{morse1953,lockwood1963}
\begin{align}
x = \frac{\sinh u}{\cosh u - \cos v}\text{ \ and \ }y = \frac{\sin
v}{\cosh u-\cos v}.
\end{align}
The domains of $u$ and $v$ for the four different regions are shown
in Fig.~\ref{fig2:structure2d}. The cut of the interface can be
expressed in terms of the angle $w$, which we, with no loss of
generality, choose between $0<w<\pi$. The surface of the cylinder is
defined such that $v=w$ on the part above the interface and
$v=w-\pi$ on the part below the interface. In terms of $w$, the
radius of the cylinder is given as $r = 1/\sin w$ and the distance
from the center of the cylinder to the interface is given as
$d=|\cot w|$. It should be noted that the geometry is defined such
that the distance unit is half the intersection of the nanowire by
the substrate plane, i.e. half the dashed line in
Fig.~\ref{fig2:structure2d}. When investigating a geometry with a
specific physical dimension the polarizabilities obtained in the
present work should therefore be multiplied by the square of this
distance measured in physical units in order to convert the results
into standard units.

The analysis is started by assuming that all four dielectric
constants are linear, isotropic, homogenous, and frequency dependent
and that the nanowire diameter is small compared to the wavelength
(e.g. $<50$~nm at optical wavelengths). The latter allows us to
perform an electrostatic analysis. We therefore introduce the
electrostatic potential $\phi(\mathbf{r})$ which is related to the
electrostatic electric field as $\mathbf{E}(\mathbf{r}) =
-\nabla\phi(\mathbf{r})$. The electrostatic potential must fulfil
Laplace's equation
\begin{align}
\nabla^2\phi(\mathbf{r})=0\ \forall \ \mathbf{r},
\end{align}
with boundary conditions $\phi_i = \phi_j$ and $\varepsilon_i\hat
n\cdot\nabla\phi_i = \varepsilon_j\hat n\cdot\nabla\phi_j$ on
\emph{S}, where the subscripts $i$ and $j$ refer to the different
domains (1,2,3, and 4) and $S$ to the boundaries. In order to derive
the polarizability we first set up expressions for the electrostatic
potential and its normal derivative in each of the four regions and
then match these using the boundary conditions above.

The total potential $\phi_i(u,v)$ is written as a sum of the
incident potential $\phi_0(u,v)$ and the scattered potential
$\varphi_i(u,v)$. We consider the two linear independent
polarizations along $y$ and $x$, vertical and horizontal,
respectively. For vertical (horizontal) polarization we look for
solutions that are even (odd) functions of $x$ and thereby $u$.
Thus, for vertical polarization we may write the potential using
cosine-transformations as
\begin{align}
\phi_i(u,v) = \int_0^\infty\bar\phi_i(\lambda,v)\cos(\lambda
u)\text{d}\lambda, \ \
\bar\phi_i(\lambda,v)=\frac{2}{\pi}\int_0^\infty\phi_i(u,v)\cos(\lambda
u)\text{d}u,\label{eq:transformvertical}
\end{align}
and for horizontal polarization we may write using
sine-transformations
\begin{align}
\phi_i(u,v) = \int_0^\infty\bar\phi_i(\lambda,v)\sin(\lambda
u)\text{d}\lambda, \ \
\bar\phi_i(\lambda,v)=\frac{2}{\pi}\int_0^\infty\phi_i(u,v)\sin(\lambda
u)\text{d}u.\label{eq:transformhorizontal}
\end{align}
In order to comply with the Laplace equation, the scattered part of
the potential in $\lambda$,$v$ coordinates may be written as
\begin{align}
\bar\varphi_i(\lambda,v)=c_i(\lambda)\cosh(\lambda
v)+s_i(\lambda)\sinh(\lambda v).
\end{align}
For vertical and horizontal polarizations the incident potentials
are given as
\begin{align}
\phi_0^{(v)}(u,v)=-\frac{\sin v}{\cosh u-\cos v}\left\{ \begin{array}{cl}1 & \text{for } v>0\\
\frac{\varepsilon_3}{\varepsilon_4}& \text{for }
v<0\end{array}\right. \ \text{ and } \ \phi_0^{(h)}(u,v) =
-\frac{\sinh u}{\cosh u-\cos v},\label{eq:incidentpotentials}
\end{align}
respectively. Utilizing the approach presented in Ref.
\cite{jung2011}, it is easy to show that the vertical and horizontal
polarizabilities may be found as
\begin{align}
\alpha^{(v)} = 4\pi\int_0^{\infty}\lambda
s_3(\lambda)\text{d}\lambda \ \text{ and } \ \alpha^{(h)} =
4\pi\int_0^{\infty}\lambda
c_3(\lambda)\text{d}\lambda.\label{eq:polarizations}
\end{align}
The constants $s_3(\lambda)$ and $c_3(\lambda)$ can be derived using
the boundary conditions. In order to utilize the boundary conditions
for the normal derivative of the potential, we need the following
two expressions for the derivative with respect to $v$ of the
incident potential. By taking the derivative of
Eq.~(\ref{eq:incidentpotentials}) and using
Eqs.~(\ref{eq:transformvertical}) and (\ref{eq:transformhorizontal})
it can be shown that
\begin{align}
\frac{\partial}{\partial v}\bar\phi_0^{(v)}(\lambda, v) =
\frac{2\lambda\cosh[\lambda(\pi-|v|)]}{\sinh[\lambda\pi]} \ \text{
and } \ \frac{\partial}{\partial v}\bar\phi_0^{(h)}(\lambda, v) =
\frac{2\lambda\sinh[\lambda(\pi-|v|)]}{\sinh(\lambda\pi)}\text{sgn}(v).
\end{align}
With the notation $C_{pq}\equiv\cosh[\lambda(p\pi-qw)]$ and
$S_{pq}\equiv\sinh[\lambda(p\pi-qw)]$ the continuity of the
potential for both polarizations yields
\begin{align}
\nonumber c_1(\lambda)C_{01}-s_1(\lambda)S_{01} &=
c_3(\lambda)C_{01}-s_3(\lambda)S_{01},\\
c_2(\lambda)C_{11}-s_2(\lambda)S_{11}&=c_4(\lambda)C_{11}-s_4(\lambda)S_{11},\label{eq:boundarypotential}\\
\nonumber c_1(\lambda)C_{10}+s_1(\lambda)
S_{10}&=c_2(\lambda)C_{10}-s_2(\lambda)S_{10},\\
\nonumber c_3(\lambda)&=c_4(\lambda).
\end{align}
For vertical polarization the boundary conditions for the normal
derivative of the potential yields
\begin{align}
\nonumber\varepsilon_1\left[-c_1(\lambda)S_{01}+s_1(\lambda)C_{01}+\frac{2
C_{11}}{S_{10}}\right] &=
\varepsilon_3 \left[-c_3(\lambda)S_{01}+s_3(\lambda)C_{01}+\frac{2C_{11}}{S_{10}}\right],\\
\varepsilon_2\left[-c_2(\lambda)S_{11}+s_2(\lambda)C_{11}+\frac{\varepsilon_3}{\varepsilon_4}\frac{2C_{01}}{
S_{10}}\right]&=
\varepsilon_4\left[-c_4(\lambda)S_{11}+s_4(\lambda)C_{11}+\frac{\varepsilon_3}{\varepsilon_4}\frac{2C_{01}}{S_{10}}\right],\label{eq:verboundaryconditionderivative}\\
\nonumber \varepsilon_1\left[c_1(\lambda)S_{10}+s_1(\lambda)
C_{10} + \frac{2}{S_{10}}\right]&=\varepsilon_2\left[-c_2(\lambda)S_{10}+s_2(\lambda)C_{10}+\frac{\varepsilon_3}{\varepsilon_4}\frac{2}{S_{10}}\right],\\
\nonumber \varepsilon_3s_3(\lambda)&=\varepsilon_4s_4(\lambda),
\end{align}
and for horizontal polarization they read
\begin{align}
\nonumber\varepsilon_1\left[-c_1(\lambda)S_{01}+s_1(\lambda)C_{01}+\frac{2
S_{11}}{S_{10}}\right] &=
\varepsilon_3 \left[-c_3(\lambda)S_{01}+s_3(\lambda)C_{01}+\frac{2S_{11}}{S_{10}}\right],\\
\varepsilon_2\left[-c_2(\lambda)S_{11}+s_2(\lambda)C_{11}+\frac{2S_{01}}{
S_{10}}\right]&=
\varepsilon_4\left[-c_4(\lambda)S_{11}+s_4(\lambda)C_{11}+\frac{2S_{01}}{S_{10}}\right],\label{eq:horboundaryconditionderivative}\\
\nonumber \varepsilon_1\left[c_1(\lambda)S_{10}+s_1(\lambda)
C_{10}\right]&=\varepsilon_2\left[-c_2(\lambda)S_{10}+s_2(\lambda)C_{10}\right],\\
\nonumber \varepsilon_3s_3(\lambda)&=\varepsilon_4s_4(\lambda).
\end{align}
By solving the equation system formed by the 8 equations of
Eqs.~(\ref{eq:boundarypotential}) and
(\ref{eq:verboundaryconditionderivative})
[Eqs.~(\ref{eq:boundarypotential}) and
(\ref{eq:horboundaryconditionderivative})], $s_3(\lambda)$
[$c_3(\lambda)$] can be found and the polarizability can be
calculated using Eq.~(\ref{eq:polarizations}). Generally both the
solution of $s_3(\lambda)$ and $c_3(\lambda)$ are fractions on the
form $2N/(DS_{10})$, where $D$ and $N$ are given as
\begin{align}
D = \sum_{p,q}D_{pq}C_{pq} \ \text{ and } N =
\sum_{p,q}N_{pq}\{C_{pq}-C_{10}\}\label{eq:DN}
\end{align}
with
\begin{align}
\nonumber D_{20}&=(\varepsilon_1+\varepsilon_2)(\varepsilon_1+\varepsilon_3)(\varepsilon_2+\varepsilon_4)(\varepsilon_3+\varepsilon_4),\\
\nonumber D_{24}&=(\varepsilon_1-\varepsilon_2)(\varepsilon_1-\varepsilon_3)(\varepsilon_2-\varepsilon_4)(\varepsilon_3-\varepsilon_4),\\
D_{22}&=2(\varepsilon_1^2-\varepsilon_2\varepsilon_3)(\varepsilon_2\varepsilon_3-\varepsilon_4^2)-2\varepsilon_1(\varepsilon_2-\varepsilon_3)^2\varepsilon_4,\\
\nonumber D_{02}&=-2(\varepsilon_1^2+\varepsilon_2\varepsilon_3)(\varepsilon_2\varepsilon_3+\varepsilon_4^2)+2\varepsilon_1(\varepsilon_2+\varepsilon_3)^2\varepsilon_4,\\
\nonumber
D_{00}&=-2(\varepsilon_1^2-\varepsilon_2\varepsilon_3)(\varepsilon_2\varepsilon_3-\varepsilon_4^2)-2\varepsilon_1(\varepsilon_2+\varepsilon_3)^2\varepsilon_4
-8\varepsilon_1\varepsilon_2\varepsilon_3\varepsilon_4.\label{eq:D_s3}
\end{align}
For $s_3(\lambda)$, i.e. vertical polarization, $N$ is computed from
\begin{align}
\nonumber N_{14} =&
-(\varepsilon_1-\varepsilon_2)(\varepsilon_1-\varepsilon_3)(\varepsilon_2-\varepsilon_4)\varepsilon_3,\\
\nonumber N_{34} =&
(\varepsilon_1-\varepsilon_2)(\varepsilon_1-\varepsilon_3)(\varepsilon_2-\varepsilon_4)\varepsilon_4,\\
N_{32} =&
(\varepsilon_1+\varepsilon_2)(\varepsilon_1-\varepsilon_3)(\varepsilon_2+\varepsilon_4)\varepsilon_4,\\
\nonumber N_{1-2} =&
(\varepsilon_1+\varepsilon_2)(\varepsilon_1+\varepsilon_3)(\varepsilon_2-\varepsilon_4)\varepsilon_3,\\
\nonumber N_{12} =&
2(\varepsilon_1^2\varepsilon_4^2+\varepsilon_2^2\varepsilon_3^2)+(\varepsilon_3+\varepsilon_4)
[\varepsilon_1^2(\varepsilon_4-\varepsilon_2)+\varepsilon_2^2(\varepsilon_3-\varepsilon_1)+(\varepsilon_1+\varepsilon_2)\varepsilon_3\varepsilon_4]\\
\nonumber
&-\varepsilon_1\varepsilon_2(\varepsilon_3^2+6\varepsilon_3\varepsilon_4+\varepsilon_4^2).\label{eq:N_s3}
\end{align}
There exists a simple symmetry between $s_3(\lambda)$ and
$c_3(\lambda)$ and thereby also $\alpha^{(v)}$ and $\alpha^{(h)}$:
\emph{By substituting $\varepsilon_i$ with $1/\varepsilon_i$ for all
$i$ $s_3(\lambda)$ transforms into $-c_3(\lambda)$ and vice versa.}
In the following, we will therefore mainly focus on vertical
polarization, because the corresponding results for horizontal
polarization can be obtained by performing simple substitutions.

The general expression for $s_3(\lambda)$ simplifies substantially
for simpler geometries. For a cut cylinder in a homogenous
surrounding ($\varepsilon_1\equiv\varepsilon$ and
$\varepsilon_2=\varepsilon_3=\varepsilon_4\equiv\varepsilon_\text{h}$)
$s_3(\lambda)$, for example, reduces to
\begin{align}
s_3(\lambda)=\frac{N_{12}(C_{12}-C_{10})+N_{32}(C_{32}-C_{10})}{D_{00}+D_{20}C_{20}+D_{02}C_{02}}\frac{2}{S_{10}},
\end{align}
where $D_{00}=-4\varepsilon\varepsilon_\text{h}$,
$D_{20}=(\varepsilon+\varepsilon_\text{h})^2$,
$D_{02}=-(\varepsilon-\varepsilon_\text{h})^2$,
$N_{12}=(\varepsilon-3\varepsilon_\text{h})(\varepsilon-\varepsilon_\text{h})/2$,
and
$N_{32}=(\varepsilon+\varepsilon_\text{h})(\varepsilon-\varepsilon_\text{h})/2$.

For $w\rightarrow0$ the geometry describes a full cylinder lying on
a substrate. In this case, the dominant contributions to the
integrals of Eq.~(\ref{eq:polarizations}) are from $\lambda\gg1$ and
we are therefore allowed to approximate as $C_{11}\approx
S_{11}\approx\exp[\lambda(\pi-w)]/2$ and $C_{10}\approx
S_{10}\approx \exp(\lambda\pi)/2$. Using this and taking
$\varepsilon_2=\varepsilon_4$ we find
\begin{align}
s_3(\lambda)\approx\frac{2(\varepsilon_1-\varepsilon_3)\varepsilon_4e^{-\lambda(2\pi+w)}[e^{2\lambda\pi}(\varepsilon_1+\varepsilon_4)-2(e^{2\lambda
w}-1)(\varepsilon_1-\varepsilon_3)]}{(\varepsilon_1+\varepsilon_4)[\varepsilon_3(\varepsilon_1+\varepsilon_4)\cosh(\lambda
w)+(\varepsilon_3^2+\varepsilon_1\varepsilon_4)\sinh(\lambda w)]},
\end{align}
which for $\lambda\gg1$ can be approximated as
\begin{align}
s_3(\lambda)\approx\frac{2(\varepsilon_1-\varepsilon_3)\varepsilon_4e^{-\lambda
w}}{\varepsilon_3(\varepsilon_1+\varepsilon_4)\cosh(\lambda
w)+(\varepsilon_3^2+\varepsilon_1\varepsilon_4)\sinh(\lambda
w)}.\label{eq:s3wsmall}
\end{align}
By performing the integral of Eq.~(\ref{eq:polarizations}) using the
above expression for $s_3(\lambda)$ we obtain the simple result for
the vertical polarizability
\begin{align}
\alpha^{(v)}=
\frac{4\pi\varepsilon_4}{w^2(\varepsilon_4-\varepsilon_3)}\text{Li}_2\left[\frac{(\varepsilon_3-\varepsilon_1)(\varepsilon_3-\varepsilon_4)}
{(\varepsilon_3+\varepsilon_1)(\varepsilon_3+\varepsilon_4)}\right],\label{eq:alpha_ver_w_small}
\end{align}
where $\text{Li}_s(z)$ is the polylogarithm (or Jonquière's function
\cite{Jonquiere1889}). A resonance at $\varepsilon_1=-\varepsilon_3$
is revealed. Note that, as expected, the polarizability is
proportional to the radius squared $r^2=1/\sin^2 w\approx1/w^2$ for
$w$ small. By performing similar approximations for the horizontally
polarized case $c_3(\lambda)$ can be found as
\begin{align}
c_3(\lambda) \approx
\frac{2(\varepsilon_1-\varepsilon_3)\varepsilon_3\exp(-\lambda
w)}{(\varepsilon_1+\varepsilon_4)\varepsilon_3\cosh(\lambda
w)+(\varepsilon_3^2+\varepsilon_1\varepsilon_4)\sinh(\lambda
w)},\label{eq:c3wsmall}
\end{align}
which integrates to
\begin{align}
\alpha^{(h)}=\frac{4\pi\varepsilon_3}{w^2(\varepsilon_4-\varepsilon_3)}\text{Li}_2\left[\frac{(\varepsilon_3-\varepsilon_1)(\varepsilon_3-\varepsilon_4)}
{(\varepsilon_3+\varepsilon_1)(\varepsilon_3+\varepsilon_4)}\right].\label{eq:alpha_hor_w_small}
\end{align}
Again a resonance at $\varepsilon_1=-\varepsilon_3$ is revealed. It
should be noted that because of the approximations utilized
$C_{11}\approx S_{11}\approx\exp[\lambda(\pi-w)]/2$ and
$C_{10}\approx S_{10}\approx \exp(\lambda\pi)/2$ the substitutional
symmetry between $s_3(\lambda)$ and $c_3(\lambda)$ has disappeared.
Eq.~(\ref{eq:c3wsmall}) cannot be obtained from
Eq.~(\ref{eq:s3wsmall}) by utilizing the simple symmetry stated
above. Our results show that, for a cylinder lying on a surface, it
is the dielectric constant of the medium above the surface that
dictates the resonances of the polarizability. In fact, the
resonance condition $\varepsilon_1 = -\varepsilon_3$ is identical to
that of a cylinder in a homogenous $\varepsilon_3$ surrounding.
However, the scaling of the polarizability for the two polarizations
are different. For vertical resonances it scales with
$\varepsilon_4$ whereas for horizontal resonances it scales with
$\varepsilon_3$.

By Taylor expanding the general expression for $s_3(\lambda)$
[Eqs.~(\ref{eq:DN}), (\ref{eq:D_s3}), and (\ref{eq:N_s3})] for
$\lambda$ small we find
\begin{align}
s_3(\lambda)\approx\frac{2}{\lambda}\left[-\frac{1}{\pi}+\frac{\varepsilon_3^{-1}+\varepsilon_4^{-1}}
{(\varepsilon_1^{-1}+\varepsilon_4^{-1})(\pi-w)+(\varepsilon_2^{-1}+\varepsilon_3^{-1})w}\right].
\end{align}
From this expansion it can be seen that $s_3(\lambda)$, and
therefore also $\alpha^{(v)}$, has a resonance at the condition
\begin{align}
(\varepsilon_1^{-1}+\varepsilon_4^{-1})(\pi-w)+(\varepsilon_2^{-1}+\varepsilon_3^{-1})w
=0.\label{eq:generalrescondition}
\end{align}
Note the dependence on $w$ of the resonance condition. For
horizontal polarization the resonance condition can be obtained by
utilizing the simple symmetry $\varepsilon_i \rightarrow
1/\varepsilon_i$. This yields
\begin{align}
(\varepsilon_1+\varepsilon_4)(\pi-w)+(\varepsilon_2+\varepsilon_3)w
=0.\label{eq:generalresconditionhor}
\end{align}

In the general case, for an arbitrary $w$, the integrals of
Eq.~(\ref{eq:polarizations}) cannot be performed analytically. It is
only possible in a few special cases. For a half-buried cylinder
$w=\pi/2$ the results are presented in Ref.~\cite{jung2011}, but
e.g. also for $w=2\pi/3$, corresponding to a 3/4 buried cylinder, an
analytical result can be obtained. We will not present these rather
comprehensive calculations here, but instead note that when the
analytical expressions of $s_3(\lambda)$ and $c_3(\lambda)$ are
given, the integrals of Eq.~(\ref{eq:polarizations}) are
straightforward to evaluate using numerical integration. There are
no problems with convergence or singularities. In fact, if the
integrals over $\lambda$ are taken from 0 to 10 a fully converged
result is obtained, except if $w$ is very small.

\section{Results}\label{sec:results}
\begin{figure}[h]
  \begin{center}
  \includegraphics[width=\textwidth]{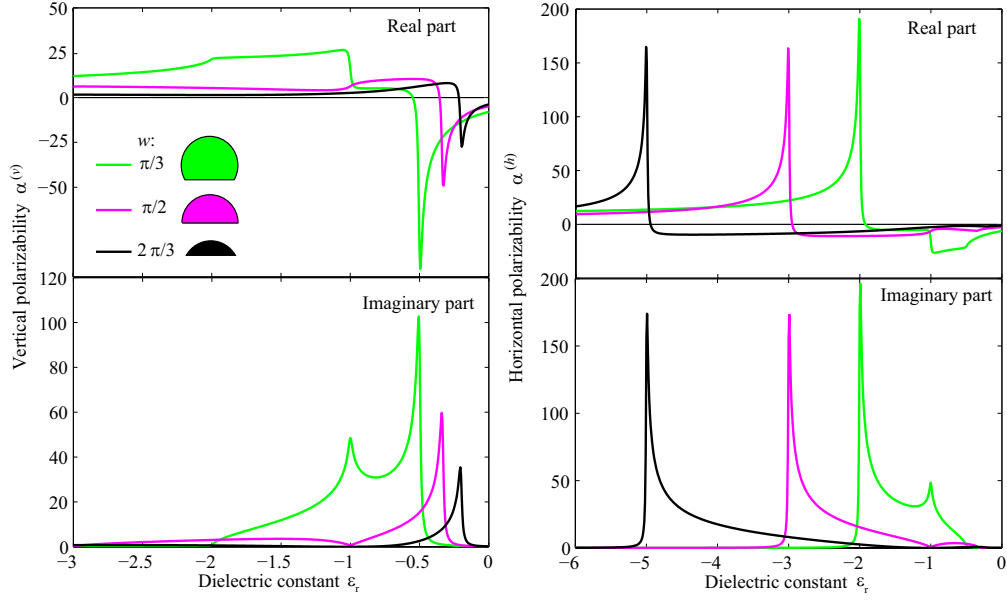}
  \caption{Polarizability as a function of $\varepsilon_\text{r}$ for three differently cut cylinders.}
  \label{fig3:conf1}
\end{center}
\end{figure}

First, we present calculations of the polarizability of three
different configurations using the approach outlined in
Sec.~\ref{sec:theory}. We consider (a) a cut cylinder in a
homogenous surrounding ($\varepsilon_1=\varepsilon$ and
$\varepsilon_2=\varepsilon_3=\varepsilon_4=\varepsilon_\text{h}=1$),
(b) a cut cylinder on a quartz surface ($\varepsilon_1 =
\varepsilon$, $\varepsilon_2 =
\varepsilon_4=\varepsilon_\text{h}=2.25$, and $\varepsilon_3=1$),
and (c) a full cylinder partly buried in a quartz substrate
($\varepsilon_1=\varepsilon_2=\varepsilon$, $\varepsilon_4 = 2.25$,
and $\varepsilon_3 = 1$). In order to identify the resonances, we
consider metal-like cylinders with complex dielectric constants that
have negative real parts and a small imaginary part. The small
imaginary part has been added to prevent singularities at the
resonances of the polarizability. Thus, the dielectric constant of
the cylinder is in the following calculations given by $\varepsilon
= \varepsilon_\text{r} + 0.01i$, where $\varepsilon_r$ is negative.
The result for configuration (a) is presented in
Fig.~\ref{fig3:conf1}.
The figures to the left display the real (top) and imaginary
(bottom) part of the vertical polarizability. The figures to the
right display the same, but for the horizontal polarizability. Three
different $w$'s are considered: $w = \pi/3$ corresponding to a 3/4
cylinder, $w = \pi/2$ a half cylinder, and $w=2\pi/3$ a 1/4
cylinder. For configuration (a) Eqs.~(\ref{eq:generalrescondition})
and~(\ref{eq:generalresconditionhor}) give the following resonance
conditions (for vertical and horizontal polarization, respectively)
\begin{align}
\varepsilon = -\varepsilon_\text{h}\frac{\pi-w}{\pi+w}\ \text{ and
}\ \varepsilon = -\varepsilon_\text{h}\frac{\pi+w}{\pi-w}.
\end{align}
For vertical polarization the three different $w$'s considered give
resonances at $\varepsilon_\text{r} = -1/2$, $-1/3$, and $-1/5$, and
for horizontal polarization we find resonances at
$\varepsilon_\text{r} = -2$, $-3$, and $-5$. All these resonances
are clearly seen both in the real and the imaginary part of the
polarizability (Fig.~\ref{fig3:conf1}). For $w$ small a resonance at
$\varepsilon_\text{r} = -\varepsilon_\text{h} = -1$ is also expected
[c.f. Eqs.~(\ref{eq:alpha_ver_w_small}) and
(\ref{eq:alpha_hor_w_small})]. For $w = \pi/3$ this resonance is
visible in the polarizability of both polarizations
(Fig.~\ref{fig3:conf1}). From Fig.~\ref{fig3:conf1} it can be seen
that the resonance in the vertical polarizability moves towards
larger $\varepsilon_r$ for an increasing angle $w$, whereas the
resonance in the horizontal polarizability moves towards smaller
$\varepsilon_r$. For ordinary plasmonic metals like silver and gold
this corresponds to a blue-shift of the resonance in the vertical
polarizability and a red-shift in the horizontal. From the presented
results it is clear that it is the geometry of the nanowire that
dictates the location of the resonance. In fact it is the ratio
between the size of the nanowire along the induced dipole moment to
the size of the nanowire perpendicular to the induced dipole moment
that is important. Thus, for the vertical polarizability it is the
height to the width ratio, which is slowly decreasing for an
increasing $w$ that accounts for the small blue-shift of the
resonance. For the horizontal polarizability it is the width to the
height ratio, which is strongly increasing with larger $w$ that
accounts for the large red-shift of the resonance.

\begin{figure}[h]
  \begin{center}
  \includegraphics[width=\textwidth]{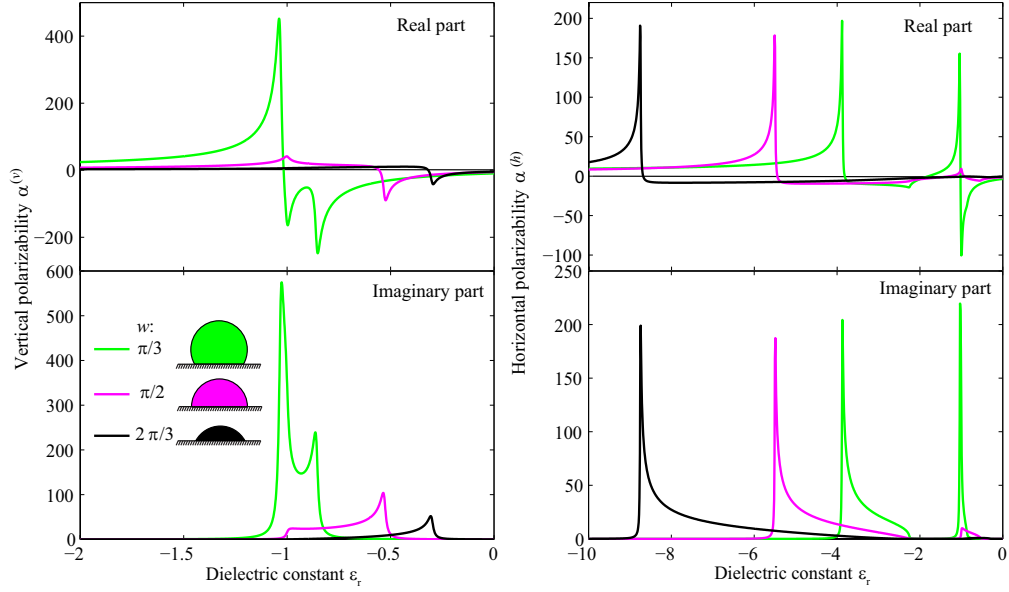}
  \caption{Polarizability as a function of $\varepsilon_\text{r}$ for three differently cut cylinders on a quartz surface.}
  \label{fig4:conf2}
\end{center}
\end{figure}

For a cut cylinder on a surface, configuration (b), the calculated
polarizability is depicted in Fig.~\ref{fig4:conf2}.
For this configuration the resonance conditions from
Eqs.~(\ref{eq:generalrescondition})
and~(\ref{eq:generalresconditionhor}) give
\begin{align}
\varepsilon =
-\varepsilon_\text{h}\varepsilon_3\frac{\pi-w}{\varepsilon_3\pi+\varepsilon_\text{h}w}\
\text{ and } \
\varepsilon=-\frac{\varepsilon_\text{h}\pi+\varepsilon_3w}{\pi-w}.
\end{align}
For the three $w$'s and vertical polarization this yields
$\varepsilon_\text{r}\approx -0.86, -0.53$, and
$\varepsilon_\text{r}=-0.3$. For horizontal polarization we expect
resonances at $\varepsilon_\text{r} = -3.875$, $-5.5$, and $-8.75$.
These resonances are all clearly seen in the calculated
polarizability (Fig.~\ref{fig4:conf2}). For configuration (b) the
resonance at $\varepsilon = -\varepsilon_3=-1$, which is expected
for $w$ small [c.f. Eqs.~(\ref{eq:alpha_ver_w_small}) and
(\ref{eq:alpha_hor_w_small})], is largely visible in both the
vertical and horizontal polarizability for $w=\pi/3$. Again it can
be seen how the resonance in the vertical (horizontal)
polarizability blue- (red)-shifts when the angle $w$ increases. As
described above, this is due to the changing geometry of the
nanowire when $w$ increases. By comparing Fig.~\ref{fig4:conf2} to
Fig.~\ref{fig3:conf1} is can be seen that all the resonances (except
for the one fixed at $\varepsilon_r =-1$ in the $w=\pi/3$
configuration) are red-shifted. This is because the effective
surrounding index that the nanowire feels in the geometry of
Fig.~\ref{fig4:conf2} is larger than in the geometry of
Fig.~\ref{fig3:conf1}. However, it should be noted that the relative
shift of the resonances is very similar in the two configurations.

\begin{figure}[h]
  \begin{center}
  \includegraphics[width=\textwidth]{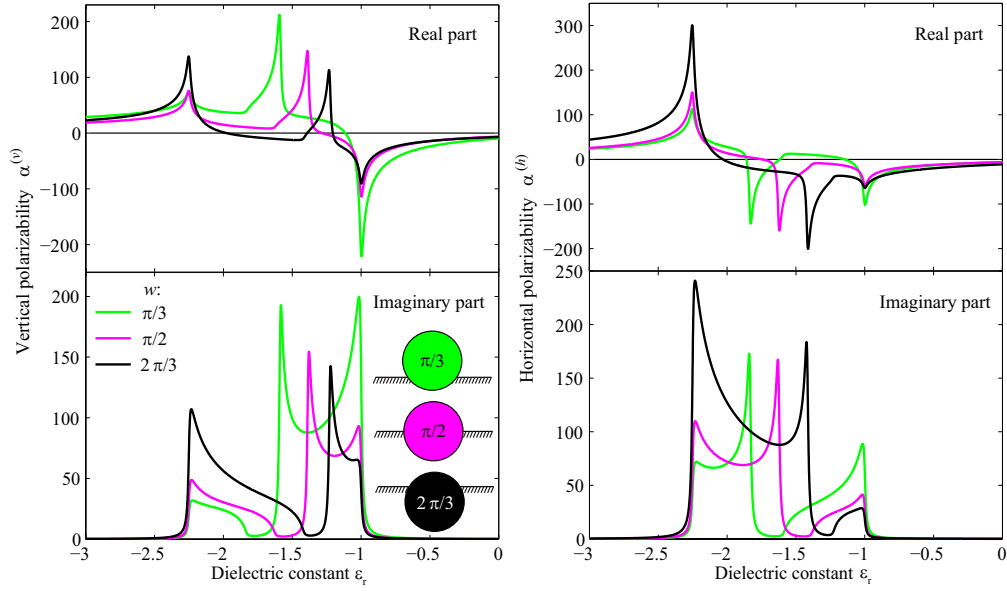}
  \caption{Polarizability as a function of $\varepsilon_\text{r}$ for three cylinders differently buried in a quartz surface.}
  \label{fig5:conf3}
\end{center}
\end{figure}

The polarizability of configuration (c), a partly buried full
cylinder, is presented in Fig.~\ref{fig5:conf3}.
In this case, Eqs.~(\ref{eq:generalrescondition})
and~(\ref{eq:generalresconditionhor}) yield
\begin{align}
\varepsilon = -
\frac{\pi\varepsilon_3\varepsilon_4}{\varepsilon_3(\pi-w)+\varepsilon_4w}\
\text{ and }\ \varepsilon =
-\frac{\varepsilon_4(\pi-w)+\varepsilon_3w}{\pi}.\label{eq:resconf3}
\end{align}
Thus, for the three $w$'s considered, we expect resonances at
$\varepsilon_\text{r} \approx -1.59$, $-1.38$, and $-1.23$ and
$\varepsilon_\text{r} \approx -1.83$, $-1.63$, and $-1.42$ for
vertical and horizontal polarization, respectively. All these
resonances are clearly seen in the polarizability
(Fig.~\ref{fig5:conf3}). However, the figure reveals that three
different resonances come into play: The one determined by
Eq.~(\ref{eq:resconf3}), and two others fixed at
$\varepsilon_\text{r} = -\varepsilon_3 = -1$ and
$\varepsilon_\text{r} = -\varepsilon_\text{h} = -2.25$. The latter
two are identical to the resonances of a full cylinder in a
homogenous surrounding described by dielectric constants
$\varepsilon_\text{h} = 1$ and $\varepsilon_\text{h} =2.25$,
respectively. It is clear from the results that the
$\varepsilon_\text{r} = -1$ resonance is strongest when $w$ is
small, i.e when a large part of the cylinder is in the medium 3,
whereas the $\varepsilon_\text{r} = -2.25$ resonance is strongest
for $w$ large, i.e. when the cylinder is largely buried in the
substrate. These observations agree nicely with the theory for $w$
very small, which predicts a resonance at $\varepsilon_1 =
-\varepsilon_3 = -1$, see Eqs.~(\ref{eq:alpha_ver_w_small}) and
(\ref{eq:alpha_hor_w_small}). From the results, in particular the
imaginary part of the polarizability, it can also be seen how the
effective index of the surrounding medium controls the cutoff
between absorption in the blue and the red part of the dielectric
spectrum. The full range of absorption is restricted to the
dielectric range $\varepsilon_r \in [-2.25,-1]$ with a gap
separating the blue and the red parts. The cutoff controlling the
position of this gap clearly reflects the degree of burial of the
nanowire. Thus, when most of the nanowire is in air, significant
absorption in the blue part of the spectrum is observed. However, as
the nanowire moves into the substrate, the large absorption shifts
towards the red part of the spectrum. The main results of the three
configurations are summarized in Table~\ref{tab:resonances}.
\begin{table}
  \centering
  \caption{Summary of the resonance conditions for the nanowire dielectric constant of the three configurations (a), (b), and (c) investigated.}
  \begin{tabular}{lllll}
    \hline
    Configuration & $w=\pi/3$ & $w=\pi/2$ & $w=2\pi/3$ & Resonance condition \\\hline
    (a) vertical & $-1/2$ & $-1/3$ & $-1/5$ & $-\varepsilon_\text{h}(\pi-w)/(\pi+w)$\\
    (a) horizontal & $-2$  & $-3$ & $-5$ & $-\varepsilon_\text{h}(\pi+w)/(\pi-w)$ \\
    (b) vertical & $-0.86$ & $-0.53$ & $-0.30$ & $-\varepsilon_\text{h}\varepsilon_3(\pi-w)/(\varepsilon_3\pi+\varepsilon_\text{h}w)$ \\
    (b) horizontal & $-3.88$ & $-5.50$ & $-8.75$ & $-(\varepsilon_\text{h}\pi+\varepsilon_3w)/(\pi-w)$ \\
    (c) vertical & $-1.59$  &$-1.38$ &$-1.23$ & $-
(\pi\varepsilon_3\varepsilon_4)/[\varepsilon_3(\pi-w)+\varepsilon_4w]$\\
    (c) horizontal & $-1.83$ & $-1.63$ & $-1.42$ & $-[\varepsilon_4(\pi-w)+\varepsilon_3w]/\pi$\\
    \hline
  \end{tabular}
  \label{tab:resonances}
\end{table}

\begin{figure}[h]
  \begin{center}
  \includegraphics[width=\textwidth]{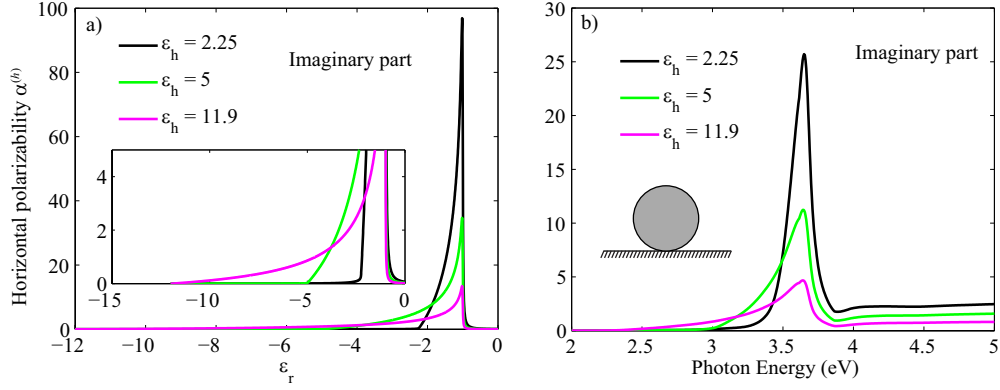}
  \caption{Imaginary part of the horizontal polarizability for a full cylinder lying on three different surfaces. (a) versus $\varepsilon_\text{r}$ and (b) versus the photon energy.
  In (a), we use $\varepsilon= \varepsilon_\text{r} +0.01i$ and in (b) we use a dielectric constant for silver taken from the experiments of Ref.~\cite{jonhson1972}.}
  \label{fig6:conf4}
\end{center}
\end{figure}

Lastly, we have calculated the polarizability in the case of
$w\rightarrow0$ taking $\varepsilon_2 =\varepsilon_4=\varepsilon_h$,
i.e the limit where the geometry resembles a full cylinder lying on
a surface. This allows us to use the approximate analytical
expressions of Eqs.~(\ref{eq:alpha_ver_w_small}) and
(\ref{eq:alpha_hor_w_small}) to calculate the polarizability. First,
we consider a cylinder with a metal-like dielectric constant
$\varepsilon = \varepsilon_r+0.01i$, where $\varepsilon_r$ is
negative. For $\varepsilon_3 = 1$ and three different substrates
$\varepsilon_\text h = 2.25$ (quartz), 5 (silicon nitride), and 11.9
(silicon), the imaginary part of the horizontal polarizability is
displayed in Fig.~\ref{fig6:conf4}~(a).
Because the polarizabilities for the two polarizations in this
configuration, except for a scaling, are identical, we only present
results for the horizontal polarizability. The results show that the
imaginary part of the polarizability rises at the resonance
$\varepsilon_\text{r} = -1$ and vanishes again for
$\varepsilon_\text{r}<-\varepsilon_\text{h}$. This is clearly seen
from the zoom [inset of Fig.~\ref{fig6:conf4}~(a)]. Thus, by
choosing a substrate with a large dielectric constant, absorption of
light in the nanowire is sustained over a relatively broad range of
wavelengths. This is more clearly seen in Fig.~\ref{fig6:conf4}~(b),
where we have calculated the imaginary part of the horizontal
polarizability versus the photon energy of a silver nanowire on
different substrates using a dielectric constant for silver taken
from the experiments of Johnson and Christy \cite{jonhson1972}. Note
how the low-frequency tail of the resonance broadens as the
dielectric constant of the substrate increases. On the
high-frequency side of the resonance the effect of interband
absorption in the silver is seen. Lastly it should be noted, that
nanocylinders on surfaces have been analyzed before using numerical
analysis \cite{nieto2001} and more recently transformation
optics~\cite{aubry2011}. Compared to the full approaches of
Refs.~\cite{nieto2001,aubry2011} the approximate polarizabilities of
Eqs.~(\ref{eq:alpha_ver_w_small}) and (\ref{eq:alpha_hor_w_small})
are easy to evaluate in that they only involve the calculation of a
single polylogarithm.

\section{Conclusion}\label{sec:conclusion}
The electrostatic polarizability of a general geometry consisting of
two nanowire segments forming a cylinder that can be arbitrarily
buried in a substrate is derived in a semi-analytical approach using
bipolar coordinates, cosine-, and sine-transformations. From the
derived expressions we have calculated the polarizability of four
important metal nanowire geometries: (1) a cut cylinder in a
homogenous surrounding, (2) a cut cylinder on a surface, (3) a full
cylinder partly buried in a substrate, and (4) a cylinder lying on a
surface. Our results give physical insight into the interplay
between the multiple resonances of the polarizability of metal
nanowire geometries at surfaces, and provide an exact, fast, and
easy scheme for optimizing metal nanowire structures for various
applications within plasmonics.

\section*{Acknowledgments} The authors gratefully acknowledge
support from the project ``Localized-surface plasmons and silicon
thin-film solar cells - PLATOS'' financed by the Villum foundation.

\end{document}